\documentclass[proceedings]{rmaa}

\usepackage{rmaacite}

\newcommand{\kms}{\,\mbox{km s$^{-1}$}}

\renewcommand{\P}[1]{%
\ifnum#1=1\hbox{OW~168--326E}\fi
\ifnum#1=2\hbox{OW~167--317}\fi
\ifnum#1=3\hbox{OW~163--317}\fi
\ifnum#1=5\hbox{OW~158--323}\fi
\ifnum#1=0\hbox{OW~171--334}\fi}

\title{Reionization of the Universe and the Photoevaporation of Cosmological
Minihalos}
\author{Paul R. Shapiro \affil{Department of Astronomy, The University of Texas at Austin, USA} \and Alejandro C. Raga \affil{Instituto de Astronom\'{\i}a, Universidad Nacional Aut\'onoma de M\'exico}}

\fulladdresses{
\item P. R. Shapiro: Department of Astronomy, University of Texas, Austin,
TX 78712, USA (shapiro@astro.as.utexas.edu)
\item A. C. Raga: Instituto de Astronom\'\i a,
  UNAM, Apartado Postal 70-264, 04510 M\'exico D. F.,
M\'exico (raga@astroscu.unam.mx)
}

\shortauthor{Shapiro \& Raga}
\shorttitle{Reionization and photoevaporation}


\resumen{Las primeras fuentes de radiaci\'on ionizante que se
  condensaron del IGM generaron frentes de ionizaci\'on en sus
  alrededores, que iluminaron a otros objetos condensados y los
  fotoevaporaron. Mostramos la retroalimentaci\'on causada por la
  reionizaci\'on del Universo en la formaci\'on de estructuras
  c\'osmicas para el caso de un mini halo de materia oscura y bariones,
  ionizado por una fuente externa con un espectro similar al de los
  cu\'asares, justo despu\'es del paso del frente de ionizaci\'on
  generado por la fuente. El mini halo es modelado, siguiendo su
  condensaci\'on del gas en expansi\'on as\'{\i} como su
  virializaci\'on, como una esfera isot\'ermica en equilibrio
  hidrost\'atico. Se presentan los resultados de la primera simulaci\'on
  hidrodin\'amica del proceso incluyendo transferencia radiativa.
  Tambi\'en se dan ejemplos de diagn\'osticos observacionales,
  incluyendo la distribuci\'on espacial del grado de ionizaci\'on en
  el flujo de C, N y O si existen trazas de estos elementos en la
  densidad columnar integrada de H~I, He~I y II, as\'{\i} como el C~IV
  a diferentes velocidades en el gas fotoevaporado, que podr\'{\i}a
  ser observado en absorci\'on en contra de la fuente de
  ionizaci\'on.}

\abstract{  
The first sources of ionizing radiation to condense out of the dark and
neutral IGM sent ionization fronts sweeping outward through their
surroundings, overtaking other condensed objects and photoevaporating them.
This feedback effect of universal reionization on cosmic structure 
formation is demonstrated here for the case of a cosmological minihalo of
dark matter and baryons exposed to an external source of ionizing radiation
with a quasar-like spectrum, just after the passage of the global 
ionization front created by the source. We model the pre-ionization
minihalo as a truncated, nonsingular isothermal sphere in hydrostatic
equilibrium following its collapse out of the expanding background universe
and virialization. Results are presented of the first gas dynamical
simulations of this process,
including radiative transfer. A sample of observational diagnostics is
also presented, including the spatially-varying ionization levels
of C, N, and O in the flow if a trace of heavy elements is present and the 
integrated column densities of H~I, He~I and II, and C~IV thru the 
photoevaporating gas at different velocities which would be measured in
absorption against a background source like that responsible for the
ionization.  }

\nonstopmode

\begin{document}

\maketitle

\section{Introduction}
\label{sec:intro}
Observations of quasar absorption spectra indicate that the universe
was reionized prior to redshift $z=5$. Recent detections of H Lyman alpha
emission lines from sources at even higher redshift ($z\leq5.6$) strengthen
this conclusion \cite{Hu98,Wey98}. As new discoveries push the observable
 horizon back in time ever-closer to the epoch of reionization at redshift 
$z>5$, increased awareness of its importance as a missing link in the 
theory of galaxy formation has caused a great renewal of interest in
universal reionization. If reionization took place early enough, then
Thompson scattering of cosmic microwave background (CMB) photons by free
electrons in an ionized intergalactic medium (IGM) would 
also have left a detectable imprint on
the CMB observed today. Recent data on the first Doppler peak in
the angular spectrum of CMB anisotropy set limits which imply a 
reionization redshift $z\lesssim40$ (model-dependent) \cite{Gri98}.
(For a review and references on reionization work prior to 1996,
the reader is referred to Shapiro 1995, while more recent developments are
summarized in Haiman \& Knox 1999.)

A review of the theory of reionization is outside the scope of this
brief report. In keeping with the focus of this meeting, I will
confine myself to a description of some recent progress on the
calculational side of the problem. To solve the full reionization
problem we must also solve the problem of how density fluctuations
led to galaxy formation and this, in turn, led to secondary energy
release in the form of ionizing and dissociating radiation by the
stars and quasars formed inside early galaxies, as well as other forms of 
energy release like supernova explosions, jets and winds, which in turn 
influenced the future course of galaxy formation. The modern context for
this description is the Cold Dark Matter (CDM) model, in which
a cold, pressure-free,
collisionless gas of dark matter dominates the
matter density and structure arises from the gravitational amplification of
a scale-free power-spectrum of
initially small-amplitude, Gaussian-random-noise primordial density
fluctuations, in a hierarchical fashion, with small mass objects collapsing
out first and merging together to form larger-mass objects which
form later. The calculation of these effects poses an enormous multi-scale 
computational challenge, involving numerical gas dynamics coupled to 
gravitational N-body dynamics, which raises the bar of cosmological 
simulation considerably by adding the requirement that radiative transfer
effects be included, as well. To simplify matters,
I will henceforth focus on just one of the central challenges of
reionization theory, the effect of cosmological ionization fronts.

\section{Ionization Fronts in the IGM} 
\label{sec:constraints}

The neutral, opaque IGM
out of which the first bound objects condensed was dramatically reheated
and reionized at some time between a redshift $z\approx50$ and $z\approx5$
by the radiation released by some of these objects.
An early analysis of the inhomogeneous nature of reionization for the case
of short-lived quasars occurring at random positions in a uniform
IGM was described by Arons \& Wingert (1972). In that treatment it was
assumed that each quasar was instantaneously surrounded by an isolated 
H~II region, a ``relict H~II region'' undergoing recombination only, each
such region filling
a volume containing just as many initially neutral H atoms as there were
ionizing photons emitted by the QSO during its lifetime. The effect
of successive generations of QSO's was accounted for on a 
statistically-averaged basis by allowing new generations of QSO's to turn 
on at random positions, including those inside pre-existing relict H~II
regions before their H atoms had fully recombined, leading eventually to 
the complete overlap of these discrete ionized zones.

When the first sources
turned on, they actually
ionized their surroundings, not instantaneously, but
rather by propagating weak, R-type
ionization fronts which moved outward supersonically with respect to both
the neutral gas ahead of and the ionized gas behind the front, racing ahead
of the hydrodynamical response of the IGM, as first described by 
Shapiro (1986) and Shapiro \& Giroux (1987). These authors solved the
problem of the time-varying radius of a spherical I-front which surrounds 
isolated sources in a cosmologically-expanding IGM analytically, taking
proper account of the I-front jump condition generalized to cosmological
conditions. They applied these solutions to determine when the I-fronts
surrounding isolated sources would grow to overlap and, thereby,
complete the reionization of the universe
(Donahue \& Shull 1987 and Meiksen \& Madau 1993
subsequently adopted a similar approach to answer that question).
The effect of density inhomogeneity on the rate of I-front propagation
was described by a mean ``clumping factor'' $c_l>1$, which
slowed the I-fronts by increasing the average recombination rate per H atom
inside clumps. This suffices to describe the rate of I-front propagation
as long as the
clumps are either not self-shielding or, if so, only
absorb a fraction of the ionizing photons emitted by the central source. 
Numerical radiative transfer methods are currently under
development to solve this problem in 3D for the inhomogeneous density 
distribution which arises as cosmic structure forms, so far limited to a 
fixed density field without gas dynamics (e.g.\ Abel, Norman, \& Madau 1999;
Razoumov \& Scott 1999). The question of what
dynamical effect the I-front had on the density inhomogeneity it
encountered, however, requires further analysis.

The answer depends on the size and density of the clumps overtaken by
the I-front.  The fate of linear density fluctuations depends upon
their Jeans number, $L_J\equiv\lambda/\lambda_J$, the wavelength in
units of the baryon Jeans length in the IGM at temperatures of order
$10^4\rm K$. Fluctuations with $L_J<1$ find their growth halted and
reversed (cf.\ Shapiro, Giroux, \& Babul 1994).  For nonlinear density
fluctuations, however, the answer is more complicated, depending upon
at least three dimensionless parameters, their internal Jeans number,
$L_J\equiv R_c/\lambda_J$, the ratio of the cloud radius $R_c$ to the
Jeans length $\lambda_J$ inside the cloud at about $10^4{\rm K}$,
their ``Str\"omgren number'' $L_s\equiv 2 R_c/\ell_s$, the ratio of
the cloud diameter $2R_c$ to the Str\"omgren length $\ell_s$ inside
the cloud (the length of a column of gas within which the unshielded
arrival rate of ionizing photons just balances the total recombination
rate), and their optical depth to H ionizing photons at 13.6 eV,
$\tau_{\rm H}$, before ionization. If $\tau_{\rm H}<1$, the I-front
sweeps across the cloud, leaving an ionized gas at higher pressure
than its surroundings, and exits before any mass motion occurs in
response, causing the cloud to blow apart. If $\tau_{\rm H}>1$ and
$L_s>1$, however, the cloud shields itself against ionizing photons,
trapping the I-front which enters the cloud, causing it to decelerate
inside the cloud to the sound speed of the ionized gas before it can
exit the other side, thereby transforming itself into a weak, D-type
front preceded by a shock. Typically, the side facing the source
expels a supersonic wind backwards towards the source, which shocks
the IGM outside the cloud, while the remaining neutral cloud material
is accelerated away from the source by the so-called ``rocket effect''
as the cloud photoevaporates (cf.\ Spitzer 1978). As long as $L_J<1$
(the case for gas bound to dark halos with virial velocity less than
$\rm 10\,km\,s^{-1}$), this photoevaporation proceeds unimpeded by
gravity. For halos with higher virial velocity, however, $L_J>1$, and
gravity competes with pressure forces. For a uniform gas of H density
$n_{\rm H,c}$, located a distance $r_{\rm Mpc}$ (in Mpc) from a UV
source emitting $N_{\rm ph,56}$ ionizing photons (in units of
$\rm10^{56}s^{-1}$), the Str\"omgren length is only
$\ell_s\cong(100\,{\rm pc}) (N_{\rm ph,56}/r_{\rm Mpc}^2)(n_{\rm
  H,c}/0.1\,\rm cm^{-3})^{-2}$.  We focus in what follows on the
self-shielded case which traps the I-front.
 
\section{The Photoevaporation of Dwarf Galaxy Minihalos
Overtaken by a Cosmological Ionization Front}
\label{sec:anal}

The importance of this photoevaporation process has long been
recognized in the study of interstellar clouds exposed to ionizing
starlight (e.g. Oort \& Spitzer 1955;
Spitzer 1978; Bertoldi 1989; Bertoldi \& McKee 1990;
Lefloch \& Lazareff 1994; Lizano et al.\ 1996). Radiation-hydrodynamical 
simulations were performed in 2D in the early 1980's of
a stellar I-front overtaking a clump inside a molecular
cloud (Sandford, Whitaker, \& Klein 1982; Klein, Sandford, \& Whitaker 1983).
More recently, 2D simulations for the case of circumstellar clouds
ionized by a single nearby star have also been performed
(Mellema et al.\ 1998). In the cosmological context, however,
the importance of this process has only recently been fully appreciated. 
In proposing the expanding minihalo model to explain Lyman alpha
forest (``LF'') quasar absorption lines, Bond, Szalay, \& Silk (1988)
discussed how gas originally confined by the gravity of dark
minihalos in the CDM model would be expelled by pressure forces
if photoionization by ionizing background radiation suddently
heated all the gas to an isothermal condition at $T\approx10^4\rm K$.
The first radiation-hydrodynamical 
simulations of the photoevaporation of a primordial density
inhomogeneity overtaken by a cosmological I-front, however,
were described in Shapiro, Raga, \& Mellema (1997, 1998).
Barkana \& Loeb (1999) subsequently considered the relative
importance of this process for dwarf galaxy minihalos of different masses
in the CDM model, using static models of uniformly-illuminated spherical
clouds in thermal and ionization equilibrium, taking H atom self-shielding
into account, and assuming that gas which is heated above the minihalo
virial temperature must be evaporated. They concluded that 50\%--90\% of 
the gas in gravitationally bound objects when reionization
occurred should have been evaporated.
\begin{figure}[h]
\parbox{0.5\textwidth}{
  \includegraphics[height=6.8cm]{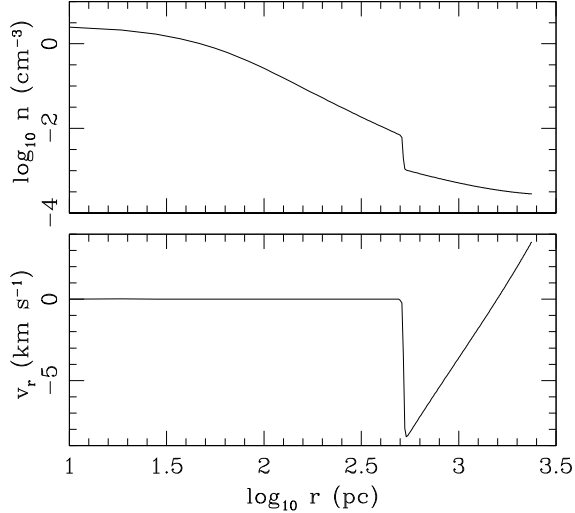}}
    \label{fig1}
\begin{minipage}{0.45\textwidth}
    \caption{MINIHALO INITIAL CONDITIONS BEFORE REIONIZATION:
Truncated, nonsingular isothermal sphere (TIS) of gas and dark matter in
hydrostatic equilibrium (Shapiro, Iliev, \& Raga 1999) surrounded by the
corresponding self-similar spherical infall for an Einstein-de Sitter
background universe (cf.\ Bertschinger 1985). (a) (Top) gas density and (b)
(Bottom) gas velocity versus distance from minihalo  center.}
\end{minipage}
\end{figure}

\begin{figure}[t]
  \parbox{0.5\textwidth}{%
    \includegraphics[height=7cm]{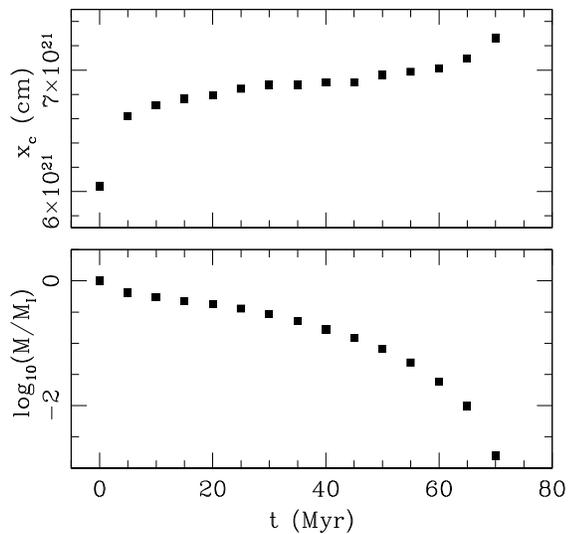}}
    \label{fig2}
\begin{minipage}{0.45\textwidth}
    \caption{IONIZATION FRONT PHOTOEVAPORATES MINIHALO:
(a) (Top) I-front position along $x$-axis versus time; (b) (Bottom) Mass
fraction of the initial mass $M_{\rm I}$ of minihalo hydrostatic 
        core which remains neutral (H~I) versus time. }
\end{minipage}
\end{figure}

\begin{figure}[h]
  \begin{center}
    \leavevmode
    \includegraphics[width=\textwidth]{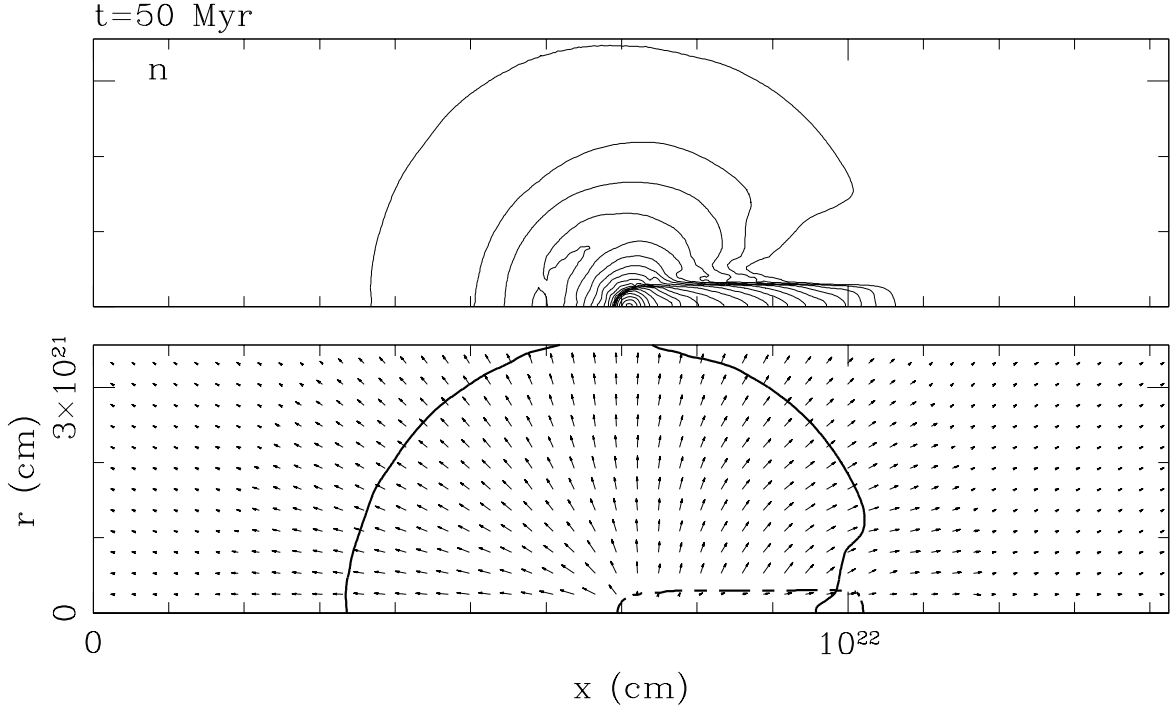}
    \caption{PHOTOEVAPORATING MINIHALO.
50 Myr after turn-on of quasar 1 Mpc to the left of computational box along
the $x$-axis. 
(a) (Upper Panel) isocontours of atomic density, logarithmically spaced, 
in $(r,x)-$plane of cylindrical coordinates; (b) (Lower Panel)
velocity arrows are plotted with length proportional to gas velocity.
An arrow of length equal to the spacing between arrows has velocity
$30 \kms$. Solid line shows current extent of gas originally in
hydrostatic core. Dashed line is I-front (50\% H-ionization contour).}
    \label{fig3}
  \end{center}
\end{figure}

\begin{figure}[p]
  \begin{center}
    \leavevmode
    \includegraphics[height=0.7\textheight]{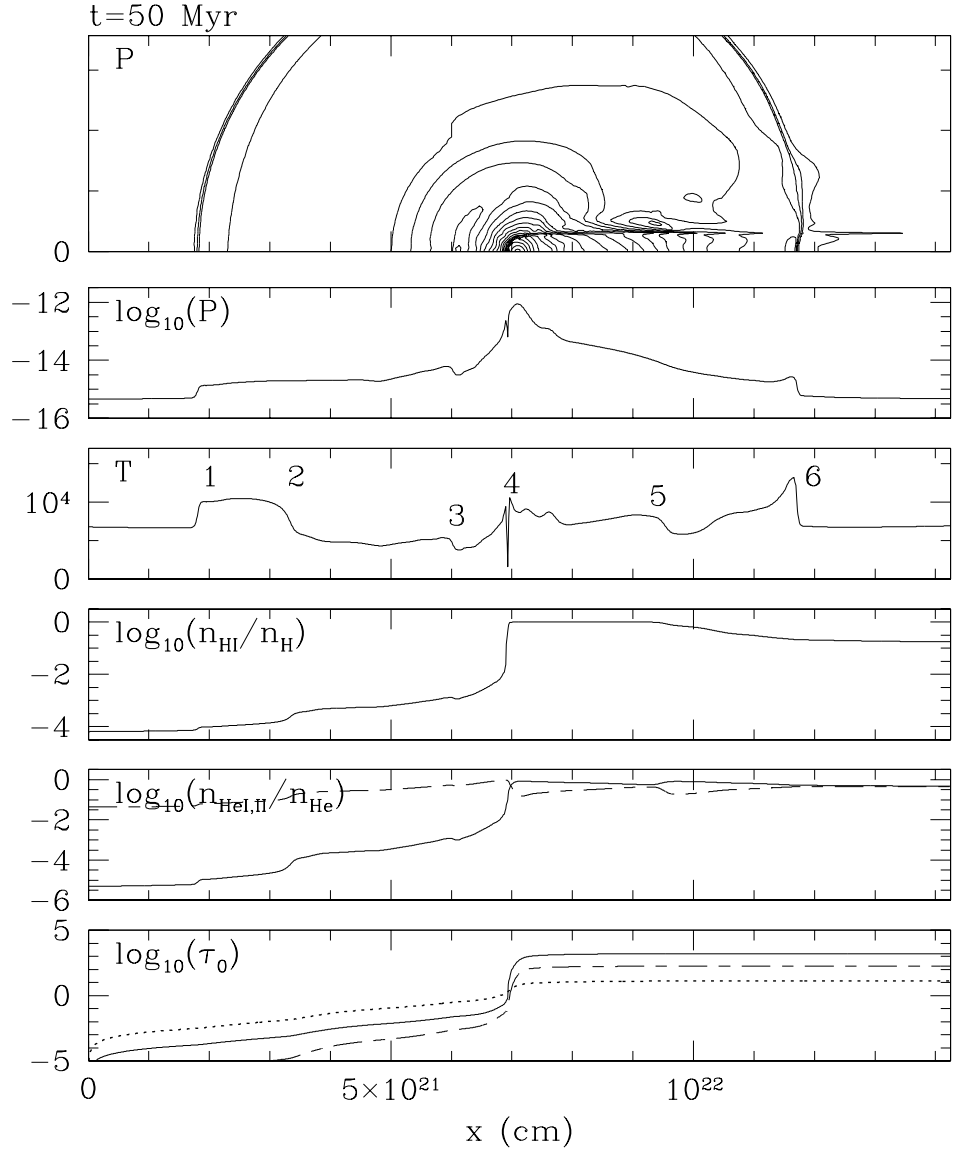}
    \caption{
      PHOTOEVAPORATING MINIHALO.  One time-slice 50 Myr after turn-on
      of quasar located 1 Mpc away from cloud to the left of
      computational box along the $x$-axis. From top to bottom: (a)
      isocontours of pressure, logarithmically spaced, in
      $(r,x)-$plane of cylindrical coordinates; (b) pressure along the
      $r=0$ symmetry axis; (c) temperature; (d) H~I fraction; (e) He~I
      (solid) and He~II (dashed) fractions; (f) bound-free optical
      depth along $r=0$ axis at the threshold ionization energies for
      H~I (solid), He~I (dashed), He~II (dotted). Key features of the
      flow are indicated by the numbers which label them on the
      temperature plot: 1 = IGM shock; 2 = contact discontinuity
      between shocked cloud wind and swept up IGM; 3 = wind shock;
      between 3 and 4 = supersonic wind; 4 = I-front; 5 = boundary of
      gas originally in hydrostatic core; 6 = shock in shadow region
      caused by compression of shadow gas by shock-heated gas outside
      shadow.}
    \label{fig4}
  \end{center}
\end{figure}

\begin{figure}
\begin{center}
\leavevmode
{\includegraphics[height=9.0cm]{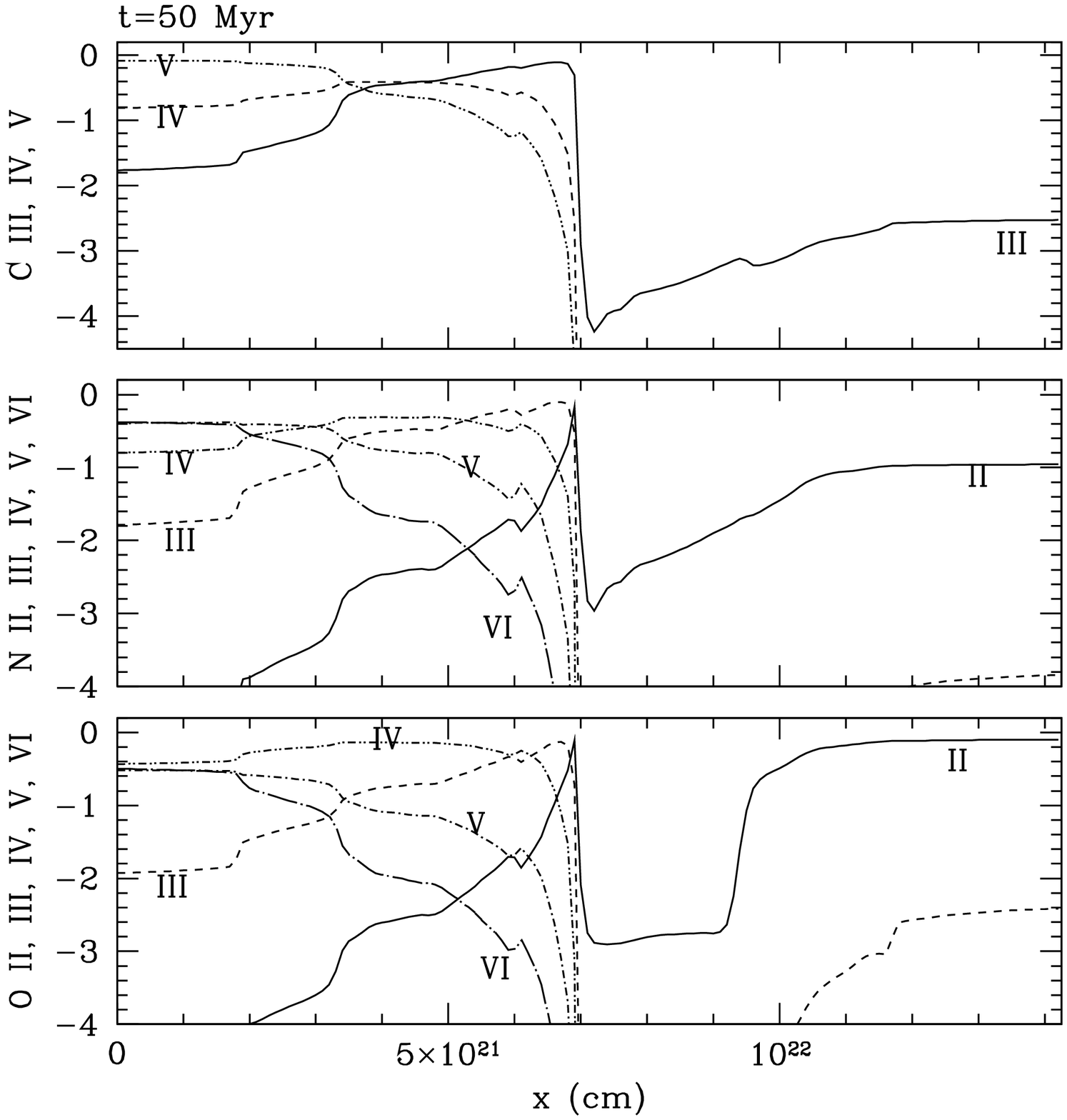}}%
    \caption{OBSERVATIONAL DIAGNOSTICS OF PHOTOEVAPORATING
        MINIHALO I: IONIZATION STRUCTURE OF METALS.
        Carbon, nitrogen, and oxygen ionic fractions along symmetry
        axis at $t = 50\rm\,Myr$.}
    \label{fig5}
\end{center}
\end{figure}

As a first study of these important effects, 
Shapiro, Raga, \& Mellema (1997, 1998) simulated the
photoevaporation of a uniform, spherical, neutral, intergalactic cloud of 
gas mass $1.5\times10^6M_\odot$, radius $R_c=0.5\,\rm kpc$, 
density $n_{\rm H,c}=0.1\,{\rm cm^{-3}}$ and $T=100\,\rm K$, located 
$1\,\rm Mpc$ from a quasar with emission spectrum $F_\nu\propto\nu^{-1.8}$
($\nu>\nu_{\rm H}$) and $N_{\rm ph}=10^{56}{\rm s}^{-1}$, initially
in pressure balance with
an ambient IGM of density $0.001\,\rm cm^{-3}$ which
at time $t=0$ had
just been photoionized by the passage of
the intergalactic R-type I-front generated when the quasar turned on
[i.e. $(L_J,L_s,\tau_H)\approx (0.1,10,10^3)$].
 [A standard top-hat 
perturbation which collapses and virializes at $z_{\rm coll}=9$,
for example, with total mass $\cong10^7M_\odot$, has circular velocity
$v_c\cong7\,\rm km\,s^{-1}$, $R_c\cong560\rm\,pc$, and
$n_{\rm H,c}=0.1\,\rm cm^{-3}$, if $\Omega_{\rm bary}h^2=0.03$ and $h=0.5$.]
The cloud contained H, He, and heavy elements at $10^{-3}$
times the solar abundance. Our simulations in 2D, axisymmetry used an
Eulerian hydro code with Adaptive Mesh Refinement and
the Van~Leer flux-splitting algorithm, which
solved nonequilibrium ionization rate equations (for H, He, C, N, O, Ne,
and S) and included an explicit treatment of radiative transfer
by taking into account the bound-free opacity of H and He 
(Raga et al.\ 1995; Mellema et al.\ 1997; Raga, Mellema, \& Lundquist 1997).
The reader is referred to Shapiro et al.\ (1997, 1998) for further details.
\begin{figure}
\begin{center}
\leavevmode
{\includegraphics[height=8cm]{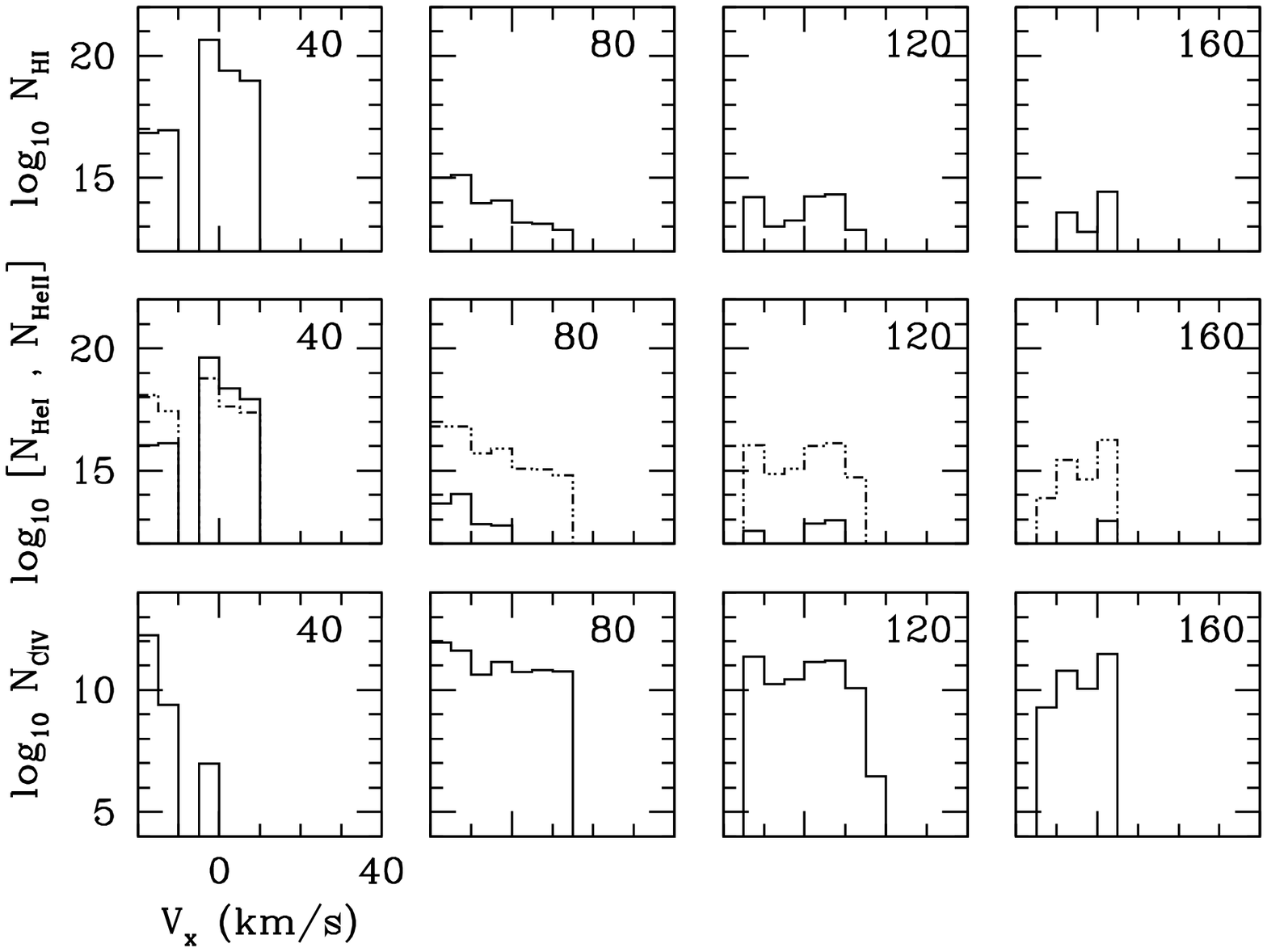}}%
    \caption{OBSERVATIONAL DIAGNOSTICS OF PHOTOEVAPORATING
        MINIHALO II. ABSORPTION LINES: 
        Cloud column densities ($\rm cm^{-2}$) along symmetry axis at
        different velocities. (Top) H~I; (Middle) He~I (solid) and
        He~II (dotted); (Bottom) C~IV. Each box labelled with time 
        (in Myrs) since QSO turn-on.}
    \label{fig6}
\end{center}
\end{figure}

Here we shall present for the first time the results of simulations
of a more realistic, cosmological minihalo, in which the uniform cloud 
described above is replaced by a self-gravitating,
centrally condensed object.
Our initial condition before ionization, shown in Figure 1, is
that of a $10^7M_\odot$ minihalo in an Einstein-de~Sitter
universe ($\Omega_{\rm CDM}=1-\Omega_{\rm bary}$; $\Omega_{\rm
bary}h^2=0.02$;
$h=0.7$) which collapses out and virializes at $z_{\rm coll}=9$,
yielding a truncated, nonsingular isothermal sphere of radius 
$R_c=0.5\,\rm kpc$ in hydrostatic equilibrium
with virial temperature $T_{\rm vir}=5900\,\rm K$
and dark matter velocity dispersion $\sigma=6.3\,\rm km\,s^{-1}$,
according to the solution of Shapiro, Iliev, \& Raga (1999),
for which the finite central density inside a radius about 1/30 of the
total size of the sphere 
is 514 times the surface density. This
hydrostatic core of radius $R_c$ is embedded in a self-similar,
spherical, cosmological infall according to Bertschinger (1985).
The results of our simulation on an $(r,x)$-grid with $256\times512$ cells
(fully refined)
are summarized in Figures~2--6.
The background IGM and infalling gas outside the minihalo are quickly
ionized, and the resulting pressure gradient in the infall region
converts the infall into an outflow.
The I-front is trapped, however, inside the hydrostatic core
of the minihalo. Figure~2(a) shows the position of the I-front inside
the minihalo as it slows from weak, R-type to weak D-type as it advances
across the original hydrostatic core. Figure~2(b) shows the mass of the
neutral zone within the original hydrostatic core shrinking as
the minihalo photoevaporates within about 70~Myrs. 
This photoevaporation time is significantly less than that found previously for a similar-mass object with
the same external source in the uniform cloud case.
Figures~3 and 4
show the structure of the photoevaporative flow 50~Myrs after the
global I-front first overtakes the minihalo, with key features of
the flow indicated by the labels on the temperature plot in Figure~4.
Figure~5 shows the spatial variation of the relative ionic abundances of
C, N, O ions along the symmetry axis after 50~Myrs.
As in the case of the uniform cloud,
Figure~5 shows the presence at 50~Myrs
of low as well as high
ionization stages of the metals.
Compared to the uniform cloud case at the same time-slice, however,
Figure~5 shows a somewhat higher degree of ionization on the
side facing the source than in those previous results.
The column densities of H~I, He~I and II, and C~IV for minihalo gas of
different velocities as seen along the symmetry axis at different times
are shown in Figure~6. 
At early times, the cloud gas resembles a weak
Damped Lyman Alpha (``DLA'') absorber with small velocity width 
($\sim10\rm\,km\,s^{-1}$) and $N_{\rm H\,I}\sim10^{20}\rm cm^{-2}$,
with a 
LF-like red wing ($\hbox{velocity width}\,\sim10\,\rm km\,s^{-1}$)
with $N_{\rm H\,I}\sim10^{16}\rm cm^{-2}$ on the side moving toward
the quasar, with a C~IV feature with
$N_{\rm C\,IV}\sim10^{12}\rm cm^{-2}$ displaced in
this same asymmetric way from the velocity of peak H~I column
density. After 160~Myr,
however, only a narrow H~I feature with LF-like column density
$N_{\rm H\,I}\sim10^{14}\rm cm^{-2}$ remains, with 
$N_{\rm He\,II}/N_{\rm H\,I}\sim10^2$ and
$N_{\rm C\,IV}/N_{\rm H\,I}\sim\rm[C]/[C]_\odot$. 
A comparison with the
results of the uniform cloud case in Shapiro et al.\ (1997, 1998)
shows that, despite their differences,
there is a surprising degree of similarity between the qualitative 
features presented there and those found here for a highly 
centrally-concentrated minihalo. Future work will extend this study to 
minihalos
of higher virial temperatures, for which gravity competes more effectively
with photoevaporation.

\acknowledgements 
This work was supported by NASA Grants NAG5-2785, NAG5-7363, and
NAG5-7821, and NSF Grant ASC-9504046,
and was made possible by a UT Dean's Fellowship and a National
Chair of Excellence, UNAM, M\'exico awarded by CONACyT in 1997 to Shapiro.
This material is based in part upon work supported by the
Texas Advanced Research Program under Grant No. 3658-0624-1999.

\enlargethispage{4ex}

\end{document}